\documentstyle[prl,aps]{revtex}

\def\Ac{\mbox{{$\cal A$}}}

\draft
\begin{document}
\input{epsf}
\twocolumn[\hsize\textwidth\columnwidth\hsize\csname@twocolumnfalse\endcsname

\title{Generalized synchronization of chaos in non-invertible maps}
\vspace{0.4in}

\author{V. Afraimovich$^{a}$, A. Cordonet$^{a}$ and  N.F. Rulkov$^{b}$}
\vspace{0.1in}

\address{
 $^{a}$ IICO-UASLP, A. Obreg\'on 64, 78000 San Luis Potos\'{\i},
SLP, M\'exico}
\address{
 $^{b}$ Institute for Nonlinear Science, University of California,
San Diego, La Jolla, CA 92093-0402}

\date{\today}
\maketitle
\begin{abstract}
The properties of functional relation between a non-invertible
chaotic drive and a response map in the regime of generalized
synchronization of chaos are studied. It is shown that despite a
very fuzzy image of the relation between the current states of the
maps, the functional relation becomes apparent when a sufficient
interval of driving trajectory is taken into account. This paper
develops a theoretical framework of such functional relation and
illustrates the main theoretical conclusions using numerical
simulations.
\end{abstract}

\narrowtext
\vskip1pc]

%%%%%%%%%%%%%%%%%
\section{Introduction}\label{sec1}
Since the first studies by Van der Pol the dynamical theory of
forced synchronization relates the synchronization phenomena to the
onset of stable response behavior of driven
oscillator~\cite{VDP,Andronov,Minorsky}. As a result of this
stability, the beats in the response oscillations disappear and a
stable periodic motion occurs in the phase space of the driven
oscillator. After that the oscillator becomes enslaved by the
periodic forcing. Two main bifurcation scenarios that lead to the
formation of the stable response behavior are the birth of stable
limit cycle on a
torus~\cite{Andronov,Arnold,Afraimovich83,Glazier88,Pikovsky2000}
and Andronov-Hopf bifurcation~\cite{Andronov,Holmes78,Mettin93}.
Based on these theoretical frameworks recent studies of
synchronization in the chaotic oscillators have lead to the
development of various notions of chaos synchronization. These
notions include: identical
synchronization~\cite{FY84,Pecora90,Rulkov96}, generalized
synchronization~\cite{AVR86,Rulkov95,Pecora95,Parlitz96}, and phase
synchronization~\cite{Pikovsky2000,Pikovsky97,Zaks99}.

An extension of the stability approach towards the case of
directionally coupled non-identical chaotic oscillators is captured
by the notion of generalized synchronization. The term, generalized
synchronization, was introduced in~\cite{Rulkov95} and used to
describe the onset of synchronization in directionally coupled
chaotic systems as the formation of a continuous mapping that
transforms a trajectory on the attractor of the drive system into a
trajectory of the response system. For the systems with invertible
dynamics this is equivalent to the formation of a continuous
mapping which links the current states of the systems when they are
settled down on the synchronous attractor.

In the case of invertible dynamics of the driving system the
relation between attracting properties of the response behavior and
the some of the properties of the  synchronization mapping has been
reviled and
proved~\cite{Hunt,Stark1997,Stark1999,Josic,Afr2000,Josic2000}. In
the case of non-invertible dynamics the relation between the
response stability and existence of the synchronization mapping was
indirectly observed with the auxiliary
method~\cite{Abarbanel96,Pyragas96}. However, the detailed and
rigorous study of this relation has not been done.

This paper presents the results of theoretical and numerical study
of the onset of functional relation between the chaotic
trajectories of driving non-invertible system and the states of the
response system. Some results about this type of synchronization
can also be found in \cite{Chubb,So2002}. To be specific we
consider the systems in the form of the following maps
%%%%%%%%%%%%%%%%
\begin{eqnarray}
 x_{n+1}&=&f(x_n),\label{mapx} \\
 y_{n+1}&=&g_c(x_n,y_n),\label{mapy}
\end{eqnarray}
%%%%%%%%%%%%%
where equation (\ref{mapx}) describes non-invertible driving system
and equation (\ref{mapy}) the response system. We assume for the
sake of definiteness, that in the system (\ref{mapx}), (\ref{mapy})
$x\in X\subset \Re^m$ and $y\in Y \subset \Re^\ell$, and that $f$
and $g_c$ are continuous. The subscript $c$ stands for a coupling
parameter. The global dynamics generated by the system
(\ref{mapx}), (\ref{mapy}) can be written in the form $F_c(x_n,y_n)
= (x_{n+1},y_{n+1})$.

%%%%%%%%%%%%%%%%%%%%%%%%%%%%%%%%%%%%%%%%%%%%%%%%%%%%%%%%%%%%%%%%%%%%%%%%%%%%%%%%%%%%
\section{Theoretical Results}\label{sec2}
%%%%%%%%%%%%%%%%%%%%%%%%%%%%%%%%%%%%%%%%%%%%%%%%%%%%%%%%%%%%%%%%%%%%%%%%%%%%%%%%%%%%%%%%%%%%%%%%%%%%%%%%%%%%%%%%%%%%%%%%%%%

The non-invetibility of $f$ implies that for any $x_0\in X$, there
are possibly infinitely many different backward orbits of $x_0$
$(x_0,x_{-1},...,x_{-n},...)$. Let us denote by $P(x_0)$ the union
of all these possible backward orbits. Denote by $p$ one of these
orbits.

Assume that there exists a partition $\{X_1,...,X_l\}$ of $X$, such
that $f|_{X_{j}}$ is one-to-one to the image\footnote{The
expression $f|_{X_{j}}$ stands for the restriction of function $f$
to the partition element $X_j$. This type of partition is standard
in the Markov maps of the interval, see for
example~\cite{kitchens}.}. We define the backward symbolic sequence
$\alpha := (\alpha_0,\alpha_1,...,\alpha_n,...)$ associated to a
backward sequence $(x_0,x_{-1},...,x_{-n},...)$ with $\alpha_i=j$
if $x_{-i}\in X_{j}$.

Since we study synchronization in dissipative systems we also
assume that there exists a ball of dissipation $B\subset
\Re^{m+\ell}$, i.e. $F_c(B)\subset Int(B)$ for any $c\in S$, where
$S$ is a region in the coupling parameter space in which system
(\ref{mapy}) has stable response behavior. Without loss of
generality we assume that $B=B_x\times B_y$, i.e. $B$ is a
rectangle, where $B_x$ (resp. $B_y$) is a ball in the $x$-space
(resp. y-space)). Denote by $\Ac_c$ the maximal attractor in $B$,
i.e. $\Ac_{c}=\cap_{n=0}^{\infty} F^n_c(B)$. Assuming that system
(\ref{mapy}) has stable response behavior, we have
%%%%%%%%%%%%%%%%%%%%%%%%%%%%%%%%%%%%%%%%%%%%%%%%%%%
\begin{equation}\label{ms}
\lim_{n\rightarrow \infty} | y_n-\tilde{y}_n|=0,
\end{equation}
%%%%%%%%%%%%%%%%%%%%%%%%%%%%%%%%%%%%%%%%%%%%%%%%%%%
where $(x_n,y_n)=F_c^n(x_0,y_0)$
($(x_n,\tilde{y}_n)=F_c^n(x_0,\tilde{y}_0)$) and $(x_0,y_0)$,
$(x_0,\tilde{y}_0)$ are arbitrary points in $B$.

Let $\Ac_{c,x}:=\Pi_x\Ac_c$ be the image of $\Ac_c$ under the
natural projection $\Pi_x$ to $X$. The set $\Ac_{c,y}$ is the
image of  $\Ac_{c}$ by $\Pi_{y}$,  the natural projection to
$Y$.

%%%%%%%%%%%%%%%%%%%%%%%%%%%%%%%%%%%%%%%%%%%%%%%%%%%%%%%%%%%%%%%%%%%%%%%%%%%%
{\bf Theorem 1}. {\it Under assumption of stable response given by
Eq.}(\ref{ms}) {\it the attractor $\Ac_c$ is the union of graphs of
infinitely many functions. Each function $h^{\alpha}$ is determined
by a symbolic backward orbit $\alpha$. Moreover each $h^{\alpha}$
is continuous.}

%%%%%%%%%%%%%%%%%%%%%%%%%%%%%%%%%%%%%%%%%%%%%%

%%%%%%%%%%%%%%%%%%%%%%%%%%%%%%%%%%%%%%%%%%%%%%
{\bf Scheme of the proof.}
%%%%%%%%%%%%%%%%%%%%%%%%%
The main point of the proof is to note that $x_0$ and $\alpha$
determine all the backward orbit $p:=(x_0,x_{-1},...,x_{-n},...)$.
From here the proof is similar to the one for the invertible case
\cite{Afr2000}. Given $p$, one can define $h^\alpha(x_0)$ as the
following limit:
 \[
h^{\alpha}(x_0):= \lim_{n\rightarrow \infty}\Pi_y F_c^n(x_{-n},y)
 \]
independently of $y\in B_y$. This limit exists because of
assumption (\ref{ms}).

Continuity is proved in the same way that in the invertible case
taking into account that $\alpha$ is the same for $x$ and
$\tilde{x}$, two close points in $X$ (see \cite{Afr2000} for
details).  $\;\Box$
%%%%%%%%%%%%%%%%%%%%%%%%%%%%%%%%%%

\vspace{5mm}

Assume that the driving system is generated by a map $f$ of the
interval $I$. Then we can say more about regularity of branches
provided that $|y_n- \tilde{y}_n|$ goes to zero exponentially fast.
Indeed the following proposition holds.

{\bf Proposition 1}.
{\it Assume that
\begin{equation}
|y_n-\tilde{y}_n|\leq A\lambda^n,
\end{equation}
where $A>0$ and $0<\lambda<1$ are constants depending on $c$.
Assume also that the one-dimensional driving system $f$ is
``hyperbolic'', i.e. there exists $n_0$ such that $(f^n)'(x)\geq
\frac{1}{\gamma_{-}}>1$ for all $n\geq n_0$ and for any $x$ for
which the derivative exists. Then each branch $h^{\alpha}$ is
Lipschitz continuous.}

%%%%%%%%%%%%%%%
{\bf Scheme of the proof.}

The proof is basically the same of the one of theorem 7
in~\cite{Afr2000}. The point is to replace the space of Lipschitz
functions $H_{L,M}$ by $H_{L,M,\alpha}$ with the restriction that
the domain for $h^\alpha\in H_{L,M,\alpha}$ consist of values of
$x$ compatible with $\alpha$.

Remark that by assumption, one has $\gamma_{-}<1$. The condition
(16) in theorem 7 in~\cite{Afr2000} can be replaced by
\begin{equation}
0<A \lambda^{n}<\frac{1}{\gamma_{-}}
\end{equation}
which  is always true if $n$ is large enough. $\; \Box$

\vspace{5mm}

Taking into account details of the proof of this proposition (which
are omitted here and can be found elsewhere~\cite{Afr2000}) one
might expect that if $f$ is not hyperbolic then continuous
synchronization functions $h^\alpha$ could be not the
Lipschitz-continuous functions. In this case the picture of
individual branches $h^\alpha$ might contain wrinkles and cups, and
as the result, might appear fuzzy in numerical simulation due to
finite resolution.

\vspace{5mm}

From now on we denote by $h$ the union of all graphs $h^{\alpha}$.
Each branch $h^\alpha$ is only defined for the values of $x$
compatible with $\alpha$. That is, if there is no backward orbit
for the value $x$ with the symbolic itinerary $\alpha$, then
$h^{\alpha}$ is not defined at $x$. Thus, in general, the number of
branches of $h$ is not necessarily the same for all $x$.

{\bf Corollary to Theorem 1} {\it If $\{X_i\}$ is a Markov
partition then for any admissible $\alpha$, the domain of
$h^\alpha$ contains an element of $\{X_i\}$, i.e. if
$\alpha=(\alpha_0,...,\alpha_n,...)$ then $D(h^\alpha)\supseteq
X_{\alpha_0}$.}

\vspace{5mm}

One may ask when two different branches of $h$ are close. The next
theorem shows that two branches that have similar recent history,
are close to each other. Branches with different recent symbols
could be either close or far from each other. As numerical
simulations discussed in Sec.\ref{sec3} we show that branches with
different recent symbols may even intersect each other.

%%%%%%%%%%%%%%%%%%%%%%%%%%%%%%%%%%
{\bf Theorem 2}. {\it Let $\{\tilde{\alpha}^i\}$ be a series of
infinitely long symbolic sequences $\tilde{\alpha}^i\in
\{1,...,l\}^\aleph$ such that $\lim_{i\rightarrow\infty}
\tilde{\alpha}^i=\alpha$ in the standard product
topology\footnote{This means in particular that the greater is $i$
the greater $N$ first symbols in $\alpha$ and $\tilde{\alpha}^i$
coincide.}~\cite{kitchens}. Then,
\[\lim_{i\rightarrow\infty}|h^{\tilde{\alpha}^i}-h^\alpha|=0 \; ,
\]
where $|h^{\beta}|:=\sup_{x\in D(\beta)}|h^\beta(x)|$}

%%%%%%%%%%%%%%%%%%%%%%%%%%%%%%%%%%%%%%%%%%%%%%
{\bf Proof.}
%%%%%%%%%%%%%%%%%%%%%%%%%

From (\ref{ms}), there exists an $N$ such that for $n\geq N$ and
for any $y_0$ and $\tilde{y}_0$
\begin{equation}\label{epsilon}
|y_n-\tilde{y}_n|\leq \epsilon
\end{equation}
This is, of course, true also for $y_0$, $\tilde{y}_0\in \Ac_{c,y}$.

Now, choose $i$ big enough in such a way that the first $N$ symbols
of $\alpha$ and $\tilde{\alpha}^i$ coincide. This means that
$(x,h^{\alpha}(x))$ and $(x,h^{\tilde{\alpha}^i}(x))$ have the same first
$N$ $x$-pre-images $(x_{-1},...,x_{-N})$. Then one can apply (\ref{epsilon})
for these points obtaining
\[
|h^{\tilde{\alpha}^i}(x)-h^\alpha(x)|\leq \epsilon\; .
\]
This is true for any $x$ compatible with $\alpha$ and $\tilde{\alpha}^i$.
The theorem is proven.
$\;\Box$

\vspace{5mm}

 If one assumes monotonicity in assumption (\ref{ms}),
that is, $|y^n-\tilde{y}^n|\leq |Y| c^n$, then a stronger result holds:
Let $\alpha$ and $\tilde{\alpha}$ have the same first $n$ symbols, then
$|h^\alpha-h^{\tilde{\alpha}}|\leq |Y|c^n$.

\vspace{5mm}

Let us remark that there are three logical possibilities for the
structure of the set $h(x)$: it could be finite, it could be
countable, and it could contain a Cantor set.

The first possibility trivially occurs in the case of identical
synchronization, that can exist even for non-invertible $f$ (see
for instance \cite{Afr2000}).

The second possibility can be justified by construction of a
special pair $f,g_c$. We believe that it could be done. It is
definitely true for the case when $f$ has zero topological entropy,
i.e. the number of admissible words $\{\alpha_0,...\alpha_{n-1}\}$
grows subexponentially as $n$ goes to infinity. For example, if the
topological Markov chain corresponding to a Markov partition
$\{X_j\}$ has zero topological entropy then the number of
admissible words grows not faster than polynomially. If
topological entropy is positive then the construction of such an
example is rather difficult, we are going to study this problem
elsewhere.

The third possibility seems to be generic in a space of pairs of
functions $f,g_c$ provided that $f$ has positive topological
entropy. It seems natural to believe that different driving signals
correspond to different outputs in the regime of synchronization.
Numerical simulations considered in Sec.\ref{sec3} confirm this
conjecture (see also Fig. 3 in \cite{Chubb} and Fig. 2d in
\cite{So2002}). However, we do not have a rigorous proof of this
statement right now.

The statements made above give a clear picture of the underlying
structure of generalized synchronization of chaos in the case when
the dynamics of drive system is non-invertible.

\section{Synchronization function in non-invertible maps: example} \label{sec3}
To illustrate the properties of generalized synchronization in
non-invertible maps consider the synchronization of logistic map
driven by tent map. In this case function $f(x_n)$ in the driving
system (\ref{mapx}) is of the form
$$ f(x)= \cases{ x/b & if $ x < b ,$ \cr
(x-1)/(b-1)  & if $ x \geq b ,$ \cr }$$ where $b$ is a control
parameter, $0<b<1$. We will consider the case when the dynamics of
the response system is given by the following map
\begin{eqnarray}
 y_{n+1}&=&(1-\epsilon)ay_n(1-y_n)+\epsilon f(x_n),\label{logy}
\end{eqnarray}
where $a$ is the control parameter of the map and $\epsilon \in
[0,1]$ is a coupling parameter. Note that upper bound of the
contraction rate in the $y$-direction denoted by $c$ in previous
section is, in this case, $c=(1-\epsilon) a$. In the numerical
simulation considered in this section the values of control
parameters are fixed $b=0.677$ and $a=3.7$.

Since the dynamics of the maps are different, the generalized
synchronization is the only possible regime of synchronization,
except to the trivial case when $\epsilon=1$. The onset of
generalized synchronization is detected with auxiliary system
approach. In this analysis we study the stability of the chaotic
response attractor in the manifold $y_n=z_n$, where variable $z_n$
is described by an exact replica of the system (\ref{logy}) which
is called auxiliary system~\cite{Abarbanel96}.

\begin{figure}
\begin{center}
\leavevmode
\hbox{%
\epsfxsize=7.5cm
\epsffile{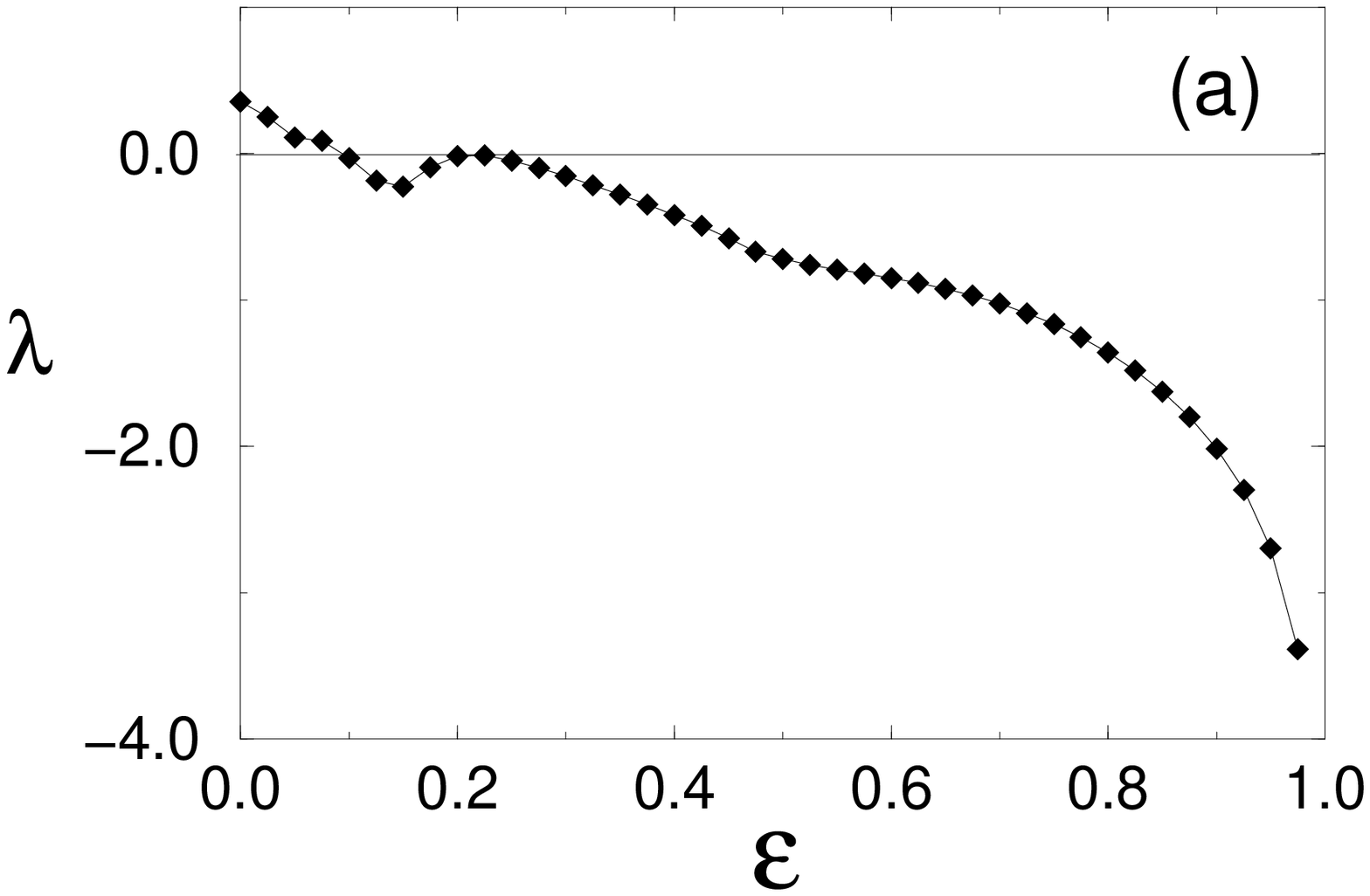}}
\hbox{%
\epsfxsize=7.5cm
\epsffile{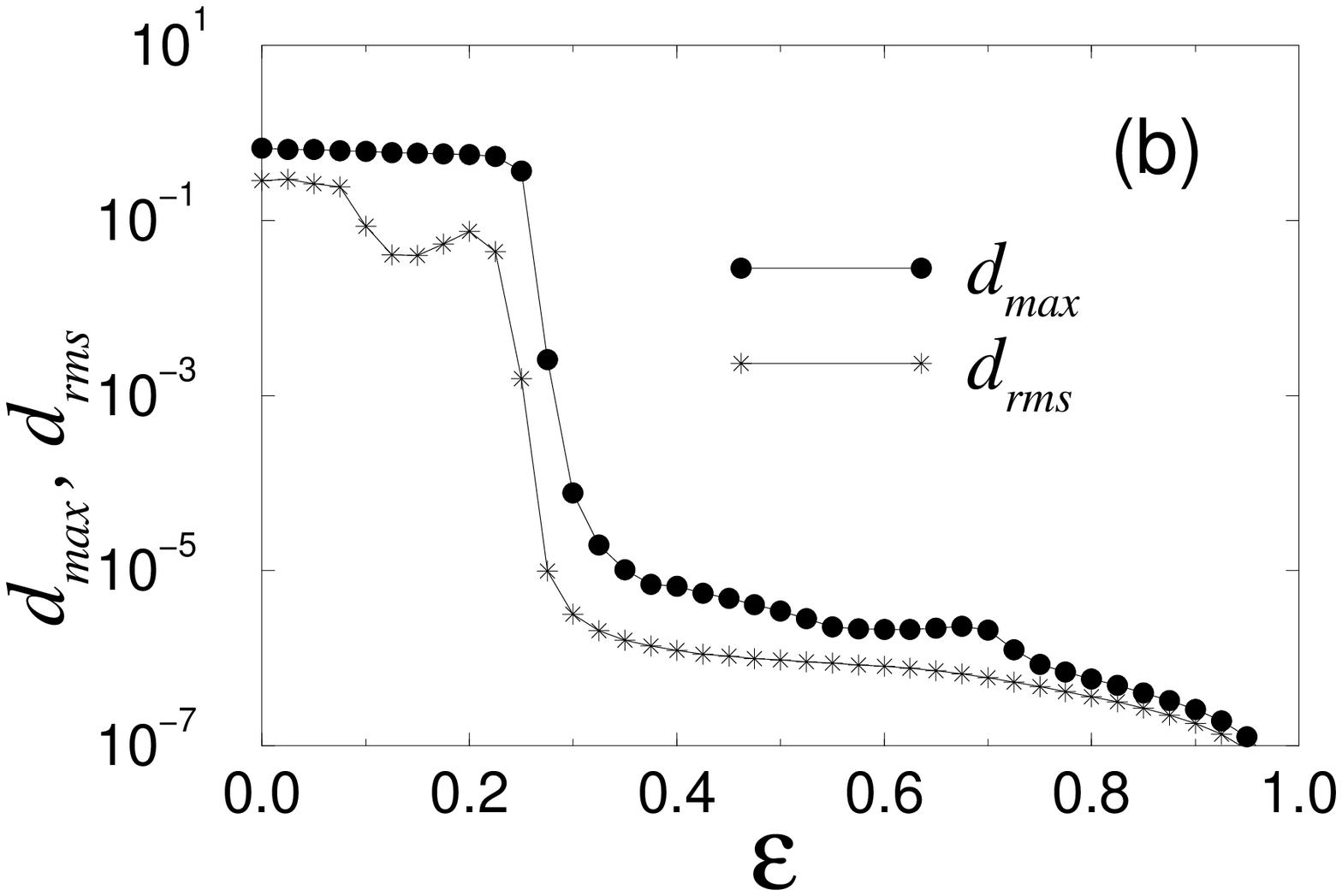}}
\end{center}
\caption{The dependance of conditional Lyapunov exponent $\lambda$
on the value of coupling parameter $\epsilon$ - (a), and the values
of maximal and rms deviations of $d_n$ computed as a function of
$\epsilon$ - (b).}
\label{fig1}
\end{figure}

It can be shown with the analysis based on contraction
mapping~\cite{neimark} that stable identical oscillations in
response and auxiliary systems are guaranteed when the values of
coupling parameter are within the interval
$1-1/a_l<\epsilon<1+1/a_r$, where $a_l=a(1-2y_n^{min})$,
$a_r=a(1-2y_n^{max})$, and $y_n^{min}$ and $y_n^{max}$ are the
leftmost and rightmost points of the attractor in the response
system. This indicates that if for the selected parameter values of
the maps the coupling is stronger than $\epsilon=0.7$, then
synchronization is monotonically stable.

More precise evaluation of the synchronization threshold can be
done with the analysis of conditional Lyapunov exponents (see,
Fig~\ref{fig1}a) and with the analysis of deviation of
response-auxiliary system from manifold $y_n=z_n$. The dependance
of maximal, $d_{max}$, and rms, $d_{rms}$, values of deviation
$d_n=z_n-y_n$ on the values of coupling parameter are presented in
Fig.~\ref{fig1}b.

\begin{figure}
\begin{center}
\leavevmode
\hbox{%
\epsfxsize=6.5cm
\epsffile{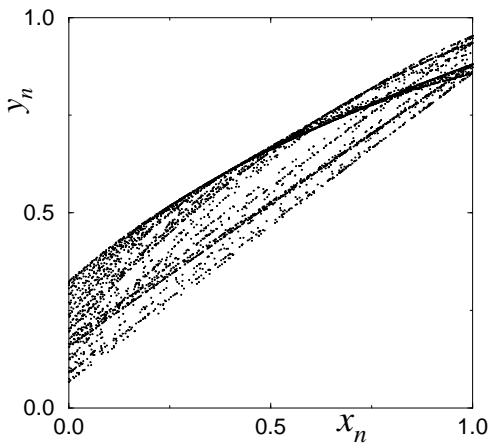}}
\end{center}
\caption{Synchronized chaotic attractor computed for $\epsilon=0.6$
plotted in the phase plane $(x_n,y_n)$. }
\label{fig2}
\end{figure}

Based upon the plots in Fig.~\ref{fig1} one would expect that the
regime of generalized synchronization takes place for the coupling
parameter values $\epsilon>0.3$ This regime assumes existence of a
continuous functional relation between the trajectories $x_n$ and
$y_n$. However, when one plots the states of the one-dimensional
phase space $y_n$ on the synchronized attractor versus the
corresponding states of one-dimensional phase space $x_n$, the
image of such relation looks fuzzy, see Fig.~\ref{fig2}. It is
clear that point to point mapping $x_n \rightarrow y_n$ cannot be
interpreted as a continuous function.

\begin{figure}
\begin{center}
\leavevmode
\hbox{%
\epsfxsize=6.5cm
\epsffile{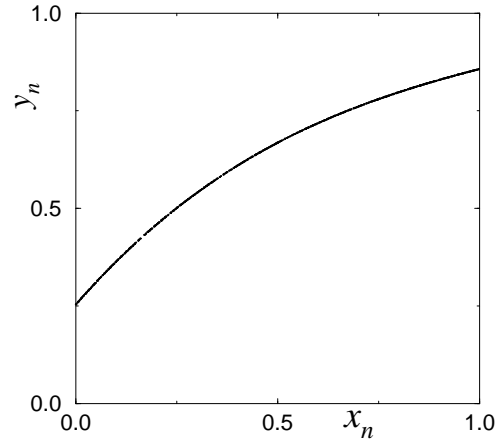}}
\end{center}
\caption{Points of the synchronized chaotic attractor shown in
Fig.~\ref{fig2} generated by the trajectories which symbolic
sequence of length $m=8$, preceding the final point $x_n$, fits to
the mask $[L,L,R,L,L,R,R,R]$. $\epsilon=0.6$}
\label{fig3}
\end{figure}

In order to unveil the synchronization function in the numerical
analysis we consider the mapping $(x_n,[\alpha_{1},...,\alpha_{m}])
\rightarrow y_n$, where $\alpha_k$ are symbolic representation of
prehistory of the trajectory $x_n$. These symbols are generated by
the tent map (\ref{mapx}) which is partitioned into two regions:
$\alpha_k=L$ if $x_{n-k}<b$ and $\alpha_k=R$ if $x_{n-k}\geq b$.
The points of the synchronous attractor whose preceding symbolic
sequence of the driving trajectory has a specific sequence (a mask)
were selected and analyzed separately. This analysis shows that as
the length of the sequence increases the cloud of points shrinks
into a curve. An example of such convergence is illustrated in
Fig.~\ref{fig3}, where the mask of 8 symbols $[L,L,R,L,L,R,R,R]$ is
studied. Similar behavior is observed for the other sequences of
the same length. In the sequences of this length the shape of the
curve varies as the symbolic sequence changes. All together these
curves form the fuzzy shape of the synchronized attractor shown in
Fig.\ref{fig2}.

\begin{figure}
\begin{center}
\leavevmode
\hbox{%
\epsfxsize=7.0cm
\epsffile{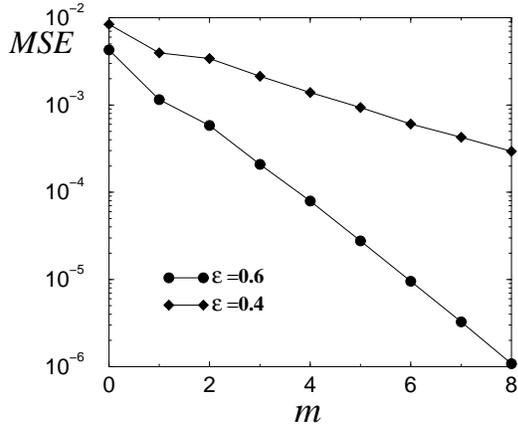}
}
\end{center}
\caption{The dependence of MSE of best polynomial fitting function
for the attractor points $y_n,x_n$ on the length $m$ of the
preceding masks $S^i_m$.}
\label{fig4}
\end{figure}

To evaluate the convergence of the image of synchronized attractor
to a continuous function $h^\alpha(x)$ we analyzed the sets of
attractor points conditioned by all possible symbolic masks
$\alpha$ of various length $m$. For each mask of preceding symbols
$S^i_m=[\alpha_{1},..,\alpha_{m}]$ we computed the best polynomial
fitting function $\phi_{S^i_m}(x)$ of order 6 using Singular Value
Decomposition (SVD) algorithm, and studied the dependance of mean
squared error (MSE), averaged over all masks of length $m$, versus
$m$. This dependence computed for the values of coupling parameter
$\epsilon=0.6$ and $\epsilon=0.4$ are shown in Fig.~\ref{fig4}.

One can see from the Fig.~\ref{fig4} that MSE decreases
exponentially fast when $m$ increases. Approximating this
dependence with exponent
\begin{eqnarray}
 MSE(m)&\sim&e^{\Lambda m},\label{mse}
\end{eqnarray}
one can find the rate of convergence, which is in the case
$\epsilon=0.6$ equals $\Lambda\approx -1.05$.

\begin{figure}
\begin{center}
\leavevmode
\hbox{%
\epsfxsize=6.0cm
\epsffile{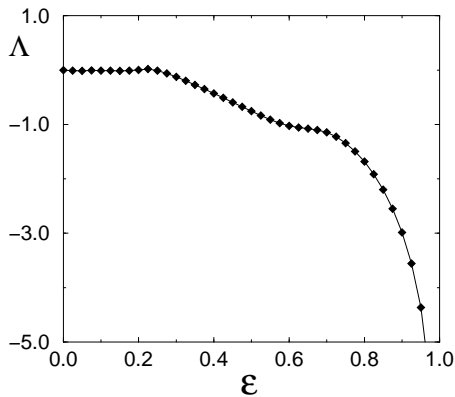}
}
\end{center}
\caption{The dependence of the convergence rate $\Lambda$
on the value of coupling parameter $\epsilon$.}
\label{fig5}
\end{figure}

Figure~\ref{fig5} shows how the convergence rate $\Lambda$
depends on the value of coupling parameter $\epsilon$. Comparing
this plot with the plot of conditional Lyapunov exponent versus
$\epsilon$ one can clearly see the similarity in these plots. This
is indicative of the fact that in the generalized synchronization
regime in our case the convergence rate $\Lambda$ is related to the
contraction rate of response system given by conditional Lyapunov
exponent $\lambda$. This relation was defined for the case of
monotonic stability of the response behavior, see Section
\ref{sec2}.

One can see from Fig.\ref{fig2} that synchronization mapping for
the current states of the drive and response system has a complex
structure of branches. The appearance of Cantor set in the
synchronized attractor caused by non-invertible driving system was
reported before in~\cite{So2002}. The formation of the fractal
structure can be explained by the analysis of deviations of the
branch caused by the change of symbol $\alpha_n$ appeared in the
symbolic sequence $n$ iterations before. For the most of the
driving trajectories, the stability of the response behavior acts
in a such way that the deviation will be reduced with the
increasing value of $n$. As the result the deviations of different
scales for different values of $n$ are responsible the formation of
the complete structure of the synchronized attractor. This explains
the formation of fractal structure in the synchronization mapping
which consists of infinite number of the branches.

\begin{figure}
\begin{center}
\leavevmode
\hbox{%
\epsfxsize=6.5cm
\epsffile{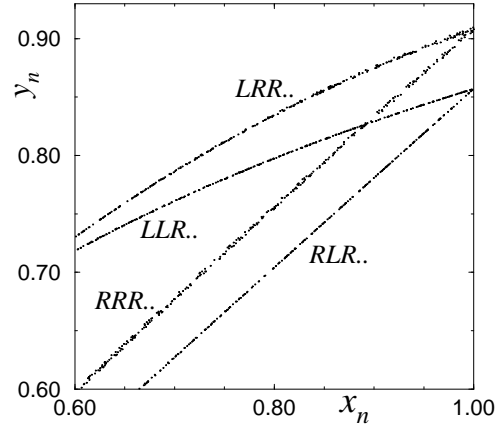}}
\end{center}
\caption{Branches of the synchronized chaotic attractor
shown in Fig.~\ref{fig2} computed for symbolic sequences of length
$m=8$, $[\alpha_1,\alpha_2,R,L,L,R,R,R]$. The values of three most
recent symbols $\alpha_1,\alpha_2,R$ are shown next to the
corresponding branch.}
\label{cs2}
\end{figure}

\begin{figure}
\begin{center}
\leavevmode
\hbox{%
\epsfxsize=6.5cm
\epsffile{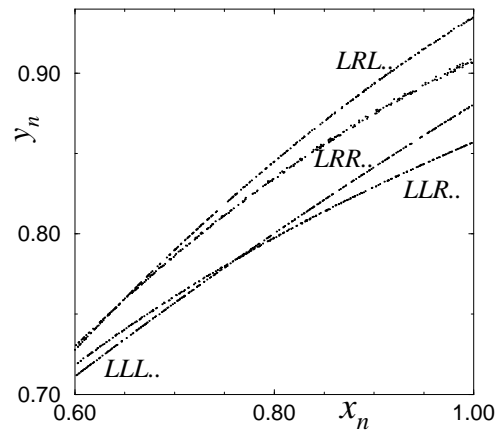}}
\end{center}
\caption{Branches of the synchronized attractor for the sequences $[L,\alpha_2,\alpha_3,L,L,R,R,R]$. The values of
three most recent symbols $L,\alpha_2,\alpha_3$ are shown next to
the corresponding branch.}
\label{cs3}
\end{figure}

This mechanics behind the formation of Cantor set is illustrated in
Figs.~\ref{cs2} and \ref{cs3}.  Figure~\ref{cs2} shows the set of
points of the synchronized attractor corresponding to the
trajectories whose 8 most recent symbols are fixed. One can see
that branches with the same most recent symbols remain relatively
close to each other, while the branches with different most recent
symbols can get apart significantly. The similar situation takes
place for the alternation of the more remote symbols, see
Fig.~\ref{cs3}. However the scale of the deviation in this case is
less then in Fig.~\ref{cs2}. Further increase of the length of
alternating symbols by one in considered mask of symbols the number
of branches doubles. The new branches appease close to the branches
with the same most recent symbols. Continuation of this process
with infinitely long mask $n \rightarrow \infty$ leads to the
formation of a fractal, Cantor type set.

\section{Conclusions}

The results of the theoretical analysis of synchronization
presented in Sec.~\ref{sec2} and the numerical analysis of
particular example considered in Sec.~\ref{sec3} allow one to draw
a number of important conclusions on typical properties of
synchronization mapping that characterize generalized
synchronization of chaos in the case of non-invertible driving
system.

The synchronization mapping, in this case, can be interpreted as a
continuous function only when all backward iterations $x_{-n}$ of
the driving trajectory $x_0$ are included in the vector of the
function arguments. We have shown that Markov partition of the
driving map can be used to describe the backward iterations in a
compact symbolic way. The set of points of synchronization
mapping plotted for the trajectories, which prehistory is
conditioned by a selected sequence of symbols, asymptotically
approach the graph of a continuous function as the length of the sequence
increases.

The set of these graphs given by all possible symbolic sequences of
infinite length form the complex fuzzy object which is typically
observed in the joint phase space of the one-dimensional chaotic
maps synchronized in the generalized sense.

An interesting feature of the considered example with 1-d maps is
that branches $h^\alpha$ remain smooth functions even when the rate
of conditional stability is rather weak. Indeed, according to the
papers~\cite{Hunt,Stark1997,Stark1999,Josic,Afr2000,Josic2000}, one
would expect that $h^\alpha$ would become non-differentiable,
H\"older continuous function as the contraction rate in the
response system becomes lower than some critical value. However,
this critical value is given by a contraction rate towards the
chaotic attractor in the driving system. In our case the driving
trajectories of the 1-d map with the specified symbolic sequence
$\alpha$ do not have contracting direction. They have only unstable
direction. As the result the contraction rate in the response
system for any given branch $h$ is always stronger than in the
driving system.

This feature is typical for the driving systems in the form of
one-dimensional hyperbolic map. In the case of non-hyperbolic 1-d
driving map function $h^\alpha$ may become non-differentiable and
contain wrinkles and cusps. Another interesting question arises
when one considers a system without exponential stability or Markov
partition? It could be expected that branches become
non-differentiable. For the moment the question is open.

\section{Acknowledgment}

This work was supported in part by a grant from the University of
California Institute for Mexico and the United States (UC MEXUS)
and the Consejo Nacional de Ciencia y Tecnologia de M\'{e}xico
(CONACYT). A.C. would like to thank the members of the INLS for
their hospitality during his visit to San Diego. A.C. is supported
by a CONACYT-ECOS-Nord contract. N.R. was supported in part by U.S.
Department of Energy (grant DE-FG03-95ER14516), the U.S. Army
Research Office (MURI grant DAAG55-98-1-0269).

\end{document}